\documentclass[doublecol]{epl2}

\title{Continuum versus discrete flux behaviour in large mesoscopic Bi$_2$Sr$_2$CaCu$_2$O$_{8+\delta}$ disks}

\author{M. R. Connolly \inst{1} \and M. V. Milo\v{s}evi\'{c} \inst{1,2} \and S. J. Bending \inst{1} \and John R. Clem \inst{3} \and T. Tamegai \inst{4}}
\shortauthor{M. R. Connolly, M. V. Milo\v{s}evi\'{c}, S. J. Bending,
John R. Clem, and T. Tamegai}

\institute{
  \inst{1} Department of Physics, University of Bath, Claverton Down, Bath, BA2 7AY,
  UK.\\
  \inst{2} Departement Fysica, Universiteit Antwerpen, Groenenborgerlaan 171, B-2020 Antwerpen,
  Belgium.\\
  \inst{3} Ames Laboratory, Department of Physics and Astronomy, Iowa State University, Ames, IA 50011-3160,
  USA.\\
  \inst{4} Department of Applied Physics, The University of Tokyo, Hongo, Bunkyo-ku, Tokyo 113-8656,
  Japan.\\
}


\abstract{Scanning Hall probe and local Hall magnetometry
measurements have been used to investigate flux distributions in
large mesoscopic superconducting disks with sizes that lie near the
crossover between the bulk and mesoscopic vortex regimes. Results
obtained by directly mapping the magnetic induction profiles of the
disks at different applied fields can be quite successfully fitted
to analytic models which assume a continuous distribution of flux in
the sample. At low fields, however, we do observe clear signatures
of the underlying discrete vortex structure and can resolve the
characteristic mesoscopic compression of vortex clusters in
increasing magnetic fields. Even at higher fields, where single
vortex resolution is lost, we are still able to track
configurational changes in the vortex patterns, since competing
vortex orders impose unmistakable signatures on ``local''
magnetisation curves as a function of the applied field. Our
observations are in excellent agreement with molecular dynamics
numerical simulations which lead us to a natural definition of the
lengthscale for the crossover between discrete and continuum
behaviours in our system.}

\begin{document}

\maketitle

Developing computational techniques that can span multiple
length-scales is a major focus of research in modern day
nanotechnology. Models of nanoscale transistors must, for example,
be able to describe them from the microscopic level of discrete
electron dynamics all the way up to the macroscopic scale of contact
pads, bond wires and packaging.  An attractive feature of using
so-called multiscale methods is that the physics of the discrete
components is embedded in parameters used at the coarse scale
without dramatically increasing the analytical and computational
overhead.  A certain degree of added complexity is often unavoidable
when using such methods, however, especially when the fine scale
properties are included without using approximations, and as a
result most physical problems are in reality still described at
either the discrete or continuum level. The application of continuum
models to systems composed of discrete components is a core concept
common to all of the natural sciences. In physics, for example, the
semi-classical conductivity tensor is a continuum approximation of
the discrete quantum-mechanical Landauer-Buttiker formalism, which
views conductance as the transmission of individual electrons.
Continuum elasticity has been hugely successful in describing many
of the properties of structural materials, yet plastic deformation
can only be understood by considering the discrete nature of solids
and the atomic-scale dislocation dynamics. Since the two approaches
are nearly always exclusive the key to successful multiscale
modelling is establishing where the crossover lengthscale between
discrete and continuum behaviours actually lies, and finding
theoretical descriptions capable of bridging it.

Vortex matter in superconductors represents an ideal model system
for studying the continuum/discrete crossover, since the vortex
density and interaction strengths can be continuously tuned by
varying the magnetic field.  One of the key topological properties
of a type II superconductor in an applied magnetic field is the
penetration and collective triangular ordering of mutually repelling
lines of magnetic flux (i.e. vortices), into the so-called Abrikosov
lattice.  If, however, the size of a superconducting sample is
reduced to mesoscopic dimensions, it is well known that the
interaction between the penetrating vortices and the circulating
edge currents can become comparable to the intervortex repulsion,
leading to the destruction of the ordered triangular lattice. Deep
in the mesoscopic regime, where the sample size is comparable to the
penetration depth $\lambda$ and/or the coherence length $\xi$, this
``topological confinement'' is so strong that the sample geometry
imposes its symmetry on multi-vortex states \cite{Schweigert1998}.
In superconducting disks this results in the arrangement of vortices
in concentric shells \cite{Baelus2004} or their collapse into a
single multiquanta vortex \cite{Kanda2004}. The behaviour of vortex
matter when the size of a sample is between the mesoscopic, where
the position of each vortex has a direct influence on its
superconducting state, and the macroscopic, where the only relevant
parameter is the ``local'' vortex density, remains an open question.
In this ``large mesoscopic'' regime, superconductivity coexists with
a large numbers of vortices, which may or may not be organized into
an Abrikosov lattice (see Ref. \cite{Cabral2004}, or \cite{Kong2003}
for the analogy with confined classical clusters.) To date only a
few studies have been able to access this regime experimentally
\cite{Grig,Hata}. Available imaging techniques have proved to be
unable to resolve individual vortices at large applied fields, while
``local'' magnetisation curves do not generally exhibit
discontinuities that can be directly associated with the transitions
between different fluxoid states seen in smaller mesoscopic disks
\cite{Geim1997}. In the absence of strong features related to
individual vortices, the quasi-continuous magnetisation curves and
flux profiles of large superconducting disks can be rather well
described by classical continuum electromagnetism \cite{Zeldov1994,
Benk1996}. The primary objective of this Letter is to address how
the discrete and continuum descriptions can be used almost
interchangeably to describe the superconducting state of disks with
dimensions near this meso-to-macro crossover point. We have explored
this regime both experimentally, using scanning Hall probe
microscopy (SHPM) \cite{Oral1996} and ``local'' magnetometry of
high-$T_c$ superconducting disks, and theoretically using modified
molecular dynamics numerical simulations. We have fabricated
periodic arrays of large mesoscopic disks on the surface of an
optimally doped Bi$_2$Sr$_2$CaCu$_2$O$_{8+\delta}$ (BSCCO) single
crystal ($T_c \approx$ 91 K). Single crystal high-$T_c$ materials
are excellent candidates for studies in the large mesoscopic regime
due to their low pinning, readily accessible temperatures of
interest, and large  $\lambda/\xi$ ratio (known as the
Ginzburg-Landau parameter $\kappa$), all of which combine to ensure
that vortices are able to attain the highly ordered ground state
shell structures.

\begin{figure}
\onefigure[width=0.6\linewidth]{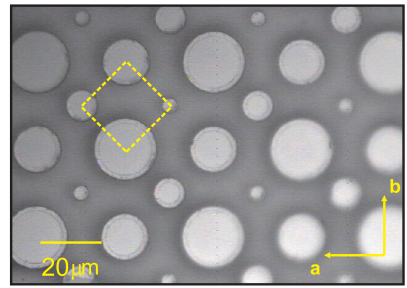} \caption{Optical micrograph of
the disk array showing the orientation of the 20 $\mu$m $\times$ 20
$\mu$m square cell (dashed lines) and the directions of the
crystallographic $a$- and $b$-axis of the BSCCO crystal.}
\label{Fig:1}
\end{figure}

\begin{figure}
\onefigure[width=\linewidth]{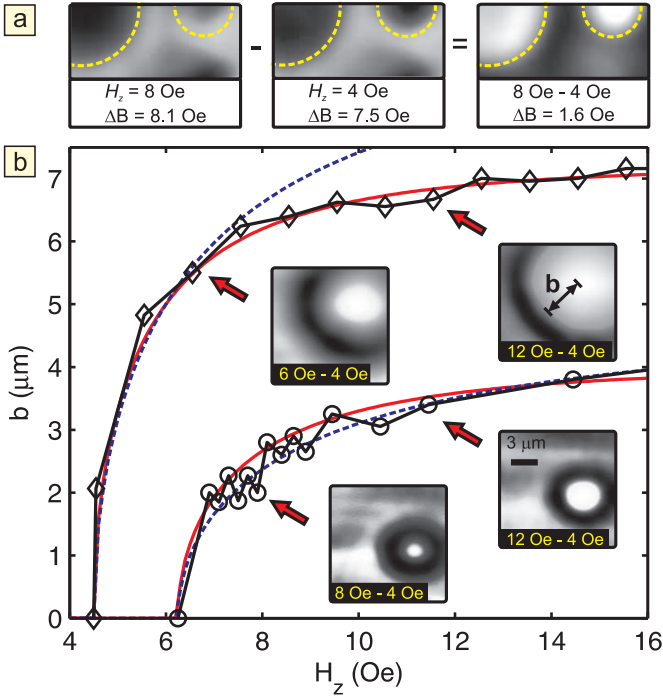} \caption{(a) 10 $\mu$m $\times$
26 $\mu$m SHPM images captured in two applied fields at $T$ = 77 K
and their difference image. Dashed outlines indicate the location of
two of the disks. (b) Radius $b$ of the vortex dome versus the
applied field, $H_z$, for the 10 $\mu$m ($\circ$) and 20 $\mu$m
($\diamond$) diameter disks. Insets show typical difference images
used to extract the experimental points. The solid lines are fits to
the measured data using a continuum model \cite{Zeldov1994} and the
dashed lines denote the relation for the instability radius of
vortex molecules from Ref. \cite{Cabral2004}. } \label{Fig:2}
\end{figure}

The disks were patterned to a depth of $d=300$ nm into the surface
of a BSCCO crystal of dimensions $\approx$ 2 mm $\times$  2 mm
$\times$ 100 $\mu$m using photolithography and Ar-ion beam milling.
The array consisted of 5 $\mu$m, 10 $\mu$m, 15 $\mu$m  and 20 $\mu$m
diameter disks sited on the corners of a square cell with 20 $\mu$m
sides inclined at 45$^\circ$ to the crystallographic $a$-axis of the
crystal (a typical region is shown in Fig. \ref{Fig:1}). The
patterned area of the crystal (typically 1-2 mm$^2$) was bonded to a
Si substrate for mechanical stability. The sample was coated with
Ti(5 nm)/Au(20 nm) films to enhance the stability of the SHPM when
in tunnelling contact. In our first experiments the array was cooled
to $T$ = 77 K and SHPM images were captured at regular (increasing)
field intervals. Fig. \ref{Fig:2}(a) illustrates how we subtract the
pixels ($B^d_{i,j} = B^2_{i,j} - B^1_{i,j}$) of two SHPM scans
captured in different magnetic fields to construct ``difference''
images.  These are used in this study to enhance the contrast due to
small numbers of penetrating vortices which cannot be resolved in
raw SHPM images. As expected \cite{Zeldov1994, Benk1996}, for fields
just above the penetration field $H_p$, a central vortex ``dome'' is
formed which expands with increasing applied field in both the 20
$\mu$m and 10 $\mu$m diameter disks [see insets in Fig.
\ref{Fig:2}(b)]. While the ``local'' edge field at which the dome
forms must be almost the same in both disks, the greater enhancement
of the applied field around the 20 $\mu$m disks (i.e. the larger
demagnetisation factor) leads to a reduction of the measured
penetration field. Figure \ref{Fig:2}(b) shows plots of the radius,
$b$, of the dome of penetrated flux as a function of the applied
field, $H_z$, for the two disk sizes. To obtain $b$($H_z$),
difference images at a given applied field were constructed with
respect to a vortex-free reference state just below penetration
($H_p$ $\approx$ 6.25 Oe and 4.5 Oe for the 10 $\mu$m and 20 $\mu$m
disks respectively). Linescans from the central peak of the dome to
the minimum at its perimeter were taken along different radial
directions to find an average value for the dome radius. The
experimental form of $b$($H_z$) illustrates how the dome expands
rapidly within a few Oersteds of the penetration field and then
saturates upon approaching the high-density screening currents close
to the disk edge.

Although the disks in our system are actually sitting on top of a
BSCCO platelet we argue that this plays no role since bulk vortex
pinning will be negligibly small at our measurement temperature.
Focussing first on the 20 $\mu$m disk, our observations find a
natural explanation within continuum models for flux penetration
\cite{Zeldov1994,Benk1996} in which vortices nucleate when the
applied field reaches the minimum field required for thermal
activation of pancake vortices over the Bean-Livingston surface
barrier (BLb).  Note that in the absence of the BLb and
demagnetisation effects this field would simply be $H_{c1}$. In
terms of the radially distributed current density, $J$($r$), the
entrance condition is equivalent to requiring a critical current
density at the sample edge. Analytic expressions for $J$($r$) follow
from the approximation that the current density is zero within the
vortex dome ($r < b$) and the magnetic induction is zero in the
vortex-free region ($b < r < R$) \cite{Zeldov1994}. The application
of the analytical results of the strip model to our disks is
justified by a number of studies, particularly the numerical
simulations of pin-free samples conducted in Ref. \cite{Brandt1999},
where the magnetisation in the two geometries shows qualitatively
identical behaviour if the strip is chosen to be twice as thick as
the disk. At a given value of $H_z$, a steady state is reached when
the radius of the dome satisfies $b = R'\sqrt{1-(H_p/H_a)^2}$  and
the solid lines in Fig. \ref{Fig:2}(b) represent fits to this
expression using $R'$ as a fitting parameter (in the case of a
strip, $R'$ is equal to the half-width). The agreement between
theory and experiment is good (we find $R'$ = 7.16 $\mu$m for the
upper curve, and 3.5 $\mu$m for the lower curve of Fig.
\ref{Fig:2}(b), close to the expected 2:1 ratio), and to our
knowledge this result is the first direct experimental validation of
the assumptions behind continuum models for the field-dependence of
the vortex dome. We note that an alternative form for the critical
entry condition was derived specifically for thin disks by Babaei
Brojeny and Clem \cite{Brojeny2003}, but was found to rise more
abruptly than the experimentally found $H_p$.
\begin{figure}
\onefigure[width=\linewidth]{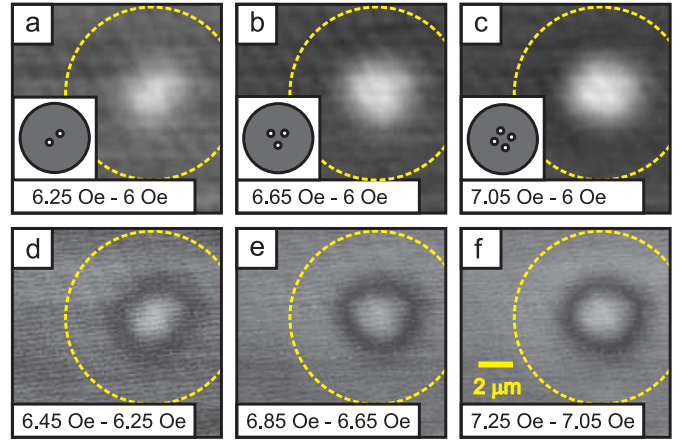} \caption{(a)-(c) Sequence of SHPM
difference images showing the evolution of vortex clusters (white)
with 2, 3, and 4 vortices in a 10 $\mu$m disk. Insets show sketches
of the proposed vortex configurations. (d)-(f) ``Difference'' images
between states with the same vorticity in different applied fields,
corresponding to the vorticity for (a)-(c) respectively.}
\label{Fig:3}
\end{figure}

\begin{figure*}
\onefigure[width=0.85\linewidth]{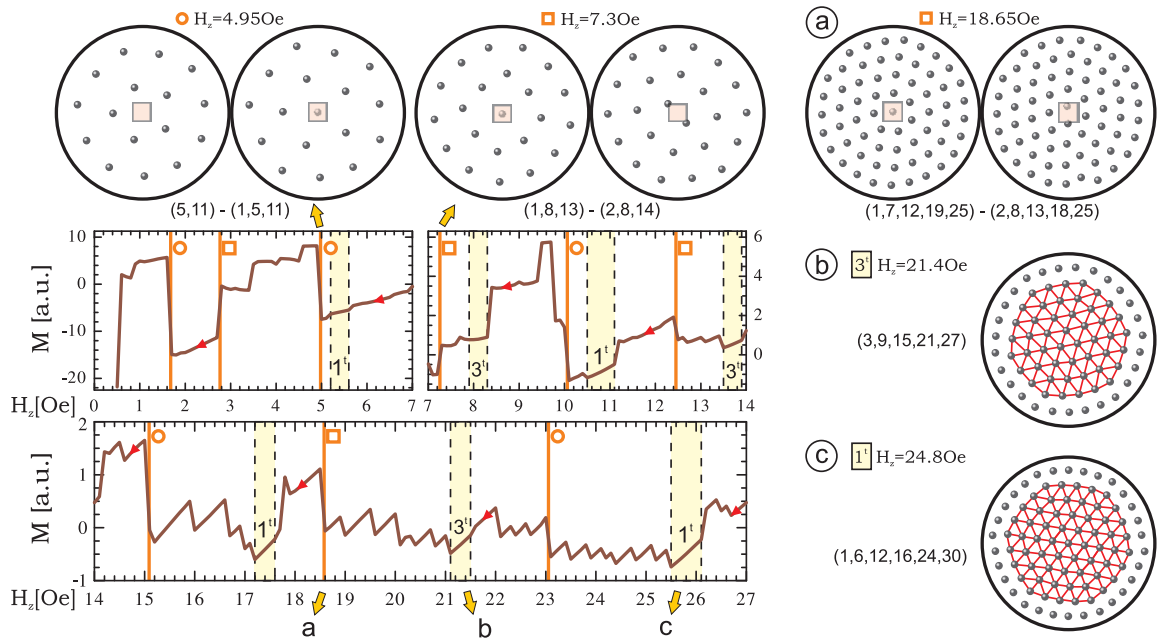} \caption{The calculated
magnetisation in decreasing applied magnetic field. The circular
snapshots show characteristic vortex configurations [for a complete
data set, see the animations in the supplementary material] and
illustrate the origin of the features in the $M$($H_z$) curve
measured across the depicted Hall probe (shaded square). In the main
panel, open circles indicate a collapse of a shell, while open
squares denote the reduction of the innermost shell to a single
vortex.} \label{Fig:4}
\end{figure*}
In contrast, it is quite apparent from Fig. \ref{Fig:2}(b) that the
continuum expression fails to describe the oscillatory behaviour of
$b$($H_z$) observed at low fields in the 10 $\mu$m disk. The origin
of these features can be seen in the SHPM images of vortex clusters
with 2, 3, and 4 vortices shown in Fig. \ref{Fig:3}(a)-(c). Although
the shape of the cluster in each case allows us to determine the
number of vortices without ambiguity, the strong overlap of the
vortex fields hampers the resolution of their exact core positions.
The field increment was reduced to just below the field step,
$\Delta H_1$, required to nucleate an additional vortex in the 10
$\mu$m disk ($\Delta H_1 = \Phi_0 / \pi R^2 \approx$ 0.263 Oe) and
the images in Fig. \ref{Fig:3}(d)-(f) show ``difference'' images
between states with the same vorticity in two different applied
fields. The dark ``rings'' correspond to a reduction in the local
flux density where the vortices have moved closer to the centre and
represent the first direct evidence for the compression of vortex
clusters in increasing applied fields. This is a distinctive
mesoscopic effect, caused by the interaction between vortices and
the increasing Meissner currents at the disk edge. Considering that
the edge currents decay away from the boundary on the scale of the
penetration depth ($\lambda \approx$ 400 nm in our sample,) this
striking observation of the ``breathing'' mode of vortex
configurations is surprising given the large size of the disk
($R>12\lambda$) and clearly supports the classification for this
disk as ``large mesoscopic''. Returning now to the discussion of
Fig. \ref{Fig:2}(b), a suitable expression for $b$($H_z$) accounting
for the discrete composition of the dome has been derived in the
limit of large $\kappa$, $\Lambda=\lambda^2/d>R$, and $d<<\lambda$
in Ref. \cite{Cabral2004}. By considering the forces on vortices
arranged in a regular polygon encircled by a ring of radius, $b$,
Cabral et al. \cite{Cabral2004} obtained $b$($H_z$) at which each
vortex cluster becomes unstable with respect to the entrance of the
next vortex. This radius is clearly the discrete counterpart of the
dome radius, $b$, in the continuum limit, and is given by $b =
R''\sqrt{1-(\Delta H_1/H_a)}$. As it stands, however, this
expression cannot be compared to the measured data because it
states, unrealistically, that penetration occurs simply when the net
flux over the area of the disk is a single flux quantum, thus
ignoring the screening effects. Since the magnitude of magnetic
screening cannot be estimated with great precision, we assume
instead that the qualitative form is correct and simply set the
offset field, $\Delta H_1$, equal to the penetration field $H_p$,
and fit the above expression to the data for both disks, again using
$R''$ as a free parameter. The best fits are shown as dashed curves
in Fig. \ref{Fig:2}(b), and were obtained for $R''$ equal to 5.1
$\mu$m and 8.9 $\mu$m for the 10 $\mu$m and 20 $\mu$m diameter
disks, respectively. The qualitative fit to the data is good for the
10 $\mu$m disk (and $R \approx R"$) but evidently fails for the 20
$\mu$m disk, strongly suggesting that the crossover from discrete to
continuum behaviour occurs between these two disk sizes.

In the light of these observations we attempt to describe the
discrete properties of our system theoretically using a modified
molecular dynamics approach based on London theory, which shows
excellent correspondence to the more complex Ginzburg-Landau
formalism in the extreme type-II limit. Ref. \cite{Cabral2004}
outlines the derivation of the vortex-vortex and vortex-boundary
interaction potentials from the expression for the free energy of
small superconducting disks. However, samples of intermediate size,
i.e. with $d\sim\lambda$ and $R>\lambda,\Lambda$ have not been
described in any simplifying limit up to date. Nevertheless,
although our samples are significantly larger than $\lambda$, the
results shown in Fig. \ref{Fig:3} suggest a very strong influence of
the boundary on the vortex configurations. On the basis of this
empirical evidence we use the mesoscopic approximation for 10 $\mu$m
diameter disks, and exploit Eqs. (12-14) from Ref. \cite{Cabral2004}
to perform the presented simulations. For the viscous drag
coefficient, $\eta$, we take the field-dependent Bardeen-Stephen
expression \cite{Bardeen1965}. A key feature of our model is the
investigation of configurational changes in the vortex states in
decreasing field. Starting from high applied field and a large
number ($\approx$ 100) of vortices, we track states with same
vorticity as long as they are stable. In decreasing field, the
previously found state is used as an initial configuration for the
next, and we allow vortices at the boundary to leave the sample as
the Bean-Livingston barrier decreases. This model closely                         
corresponds to the experimental situation in which an applied
magnetic field is swept down, since the weakening of the screening
currents close to the vortex expulsion field favours the
applicability of the London approximation (in general, London theory
neglects the suppression of superconductivity on the disk edges due
to circulating Meissner currents). For sequential vortex states in
decreasing field, the magnetisation of the sample was calculated by
integrating all (Pearl) vortex fields \cite{Pearl1964}. To link the
approach to the current experiments even more closely, we perform
the integration at a scanning height of $z=300$ nm through a
1$\times$1 $\mu$m$^2$ square area above the centre of the disk (a
situation which corresponds to a realistic Hall probe). The magnetic
response, starting from a maximum vorticity of 100, is shown as a
function of applied field in Fig. \ref{Fig:4}. As anticipated, each
change in vorticity is accompanied by a jump in the magnetisation
curve. However, in addition to these features we also observe a
series of significantly larger jumps, which stem directly from the
reconfiguration of vortices in the sample\footnote{An animated
sequence of all vortex configurations corresponding to Fig.
\ref{Fig:4} is available at http://www.ua.ac.be/milorad.milosevic
.}.

It is already well known that vortices in mesoscopic disks tend to
form shell structures \cite{Baelus2004}. Although the evolution of
the number of vortices occupying each shell strongly depends on the
size of the disk and the applied field, in decreasing fields it is
always the case that the innermost shell collapses first. Prior to
this the number of vortices in the centre of the disk gradually
decreases, and the final collapse of the shell is marked by the
integration of the last central vortex into the other shells (for
illustration, see configurations (1, 5, 11) $\rightarrow$ (5, 11) in
Fig. \ref{Fig:4}). For a Hall probe placed above the centre of the
disk, there is consequently always a sharp increase in the measured
magnetisation when a shell collapses, as is indicated by open
circles in Fig. \ref{Fig:4} (the first such transition is at 23.1
Oe). In agreement with studies of colloidal systems \cite{bedan}, we
never observe more than 5 vortices in the new innermost shell.
Further decrease of the applied field relieves the magnetic pressure
at the centre, and the number of vortices in the central cluster
decreases. The effect of the transitions between multi-vortex
structures is subtle because it depends on the size of the cluster
compared to the size of the Hall probe. In principle, the reduction
of the vorticity under the probe increases the measured
magnetisation (see for instance the transition at 18.65 Oe and the
corresponding configurations in Fig. \ref{Fig:4}). However, for low
vortex densities the cluster size may exceed the size of the probe.
With all vortices in a cluster being effectively around the probe,
the measured magnetisation now decreases when the vorticity is
lowered (see transitions marked by an open square for fields below
10 Oe, and inset configurations for $H_z = 7.3$ Oe in Fig.
\ref{Fig:4}).

In addition to the discontinuities in the magnetisation that are due
to reconfiguration, we have also identified several vortex states of
pronounced stability, shown by dashed lines in Fig. \ref{Fig:4}. All
of these states show a triangular arrangement of vortices (with
either one (1$^t$) or three (3$^t$) vortices in the disk centre)
which gradually deforms to a shell structure near the disk boundary.
{\it In actual fact, the latter structures are simultaneously
triangular and shell-like, and therefore clearly minimise the energy
of a large mesoscopic disk by combining the energetically favourable
vortex arrangements seen in the bulk and mesoscopic limits}. This
absolute minimisation of intervortex interactions results in a
stable locking of the vorticity over a finite range of applied
fields, as observed in Fig. \ref{Fig:4}.
\begin{figure}
\onefigure[width=0.8\linewidth]{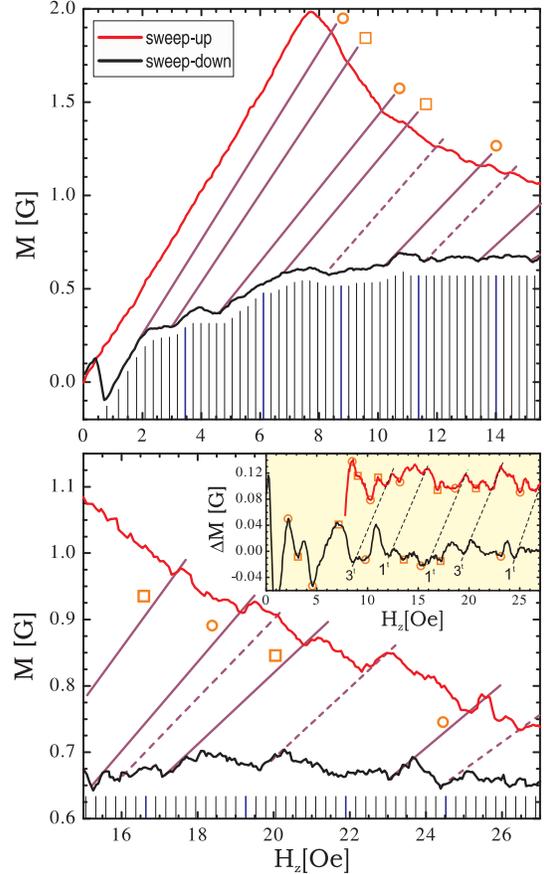} \caption{Experimentally
obtained ``local'' $M$($H_z$) curves, for increasing (red) and
decreasing (black) fields. Slanted lines indicate vortex transitions
of the same kind in sweep-up and sweep-down traces; connected points
on both curves are clearly marked in the inset, where the background
variation of $M$($H_z$) has been subtracted to extract the
discontinuities arising from configurational changes. Vertical lines
show the expected ground state transition fields between states with
sequential vorticities \cite{Baelus2004}, where the blue lines
indicate the fields at which the vorticity is a multiple of ten to
facilitate the counting.} \label{Fig:5}
\end{figure}

In Fig. \ref{Fig:5} we present the experimentally measured ``local''
magnetisation ($M = B_z - \mu_0 H_z$) captured with the SHPM Hall
sensor parked at the centre of a 10 $\mu$m disk. The applied field,
$H_z$, was swept around a measurement cycle between $\pm$ 36 Oe and
the local magnetic induction at the Hall sensor, $B_z$, was recorded
at each point. The first ``virgin'' trace was discarded, and the
subsequent 50 cycles then averaged to improve the signal-to-noise
ratio still further. The top curve in Fig. \ref{Fig:5} was obtained
in increasing field and is different to the curve measured in
decreasing field due to hysteretic flux entry/exit.  Nevertheless,
by following the pattern of magnetic signatures shown in Fig.
\ref{Fig:4}, we were able to identify and link pronounced features
in sweep-up and sweep-down curves, as labelled by the slanted lines
in Fig. 5. Remarkably, almost all magnetisation jumps predicted by
the numerical simulation were indeed found in the experimental data
at virtually the same values of the applied field on the sweep-down
leg of the curve. The solid slanted lines in Fig. \ref{Fig:5}
therefore represent transitions of the same type: either a new shell
formation or a single- to multi-vortex transition in the innermost
shell. We emphasise that while the vortex configurations at which
these transitions occur are not necessarily the same in increasing
and decreasing fields, the states with the same vorticity in the
combined shell-triangular structure, which are linked by the dashed
lines, do appear in both sweep directions due to their enhanced
topological stability. Regardless of their exact origin, it is clear
that the slopes of the transition lines drawn in Fig. \ref{Fig:5}
gradually decrease with increasing field. At larger fields, and
consequently larger vorticity, the screening currents are strongly
compensated by the vortex currents, and the lines gradually deviate
from the Meissner slope observed prior to flux penetration. For the
sake of completeness we also show the expected vorticity in the
ground-state as a function of applied field in Fig. 5, using the
expression from Ref. \cite{Baelus2004}.

Finally, we use this competition between the triangular ordering of
a lattice and deformation into vortex shells to define a criterion
for the meso- to macroscopic crossover. For high vortex densities, a
perfect triangular lattice would be ideally fitted into a
superconducting disk in such a way that apices of its hexagonal
footprint sit at a distance $\lambda$ from the disk boundary.
Vortices along the straight edges of the outer hexagon experience a
net outward force which makes the formation of a shell more
favourable and diminishes with increasing size of the disk.
Balancing all the forces acting on the central vortex on one edge of
this outer hexagon, we obtain an equality
\begin{eqnarray}
\label{Eq:MesoMacro}
\frac{R-\Lambda}{a}\sqrt{3}f_{vv}(r,r-\frac{a}{R})+\frac{r}{1-r^2}
=rL\frac{2.491+0.617L}{0.402+0.598L},
\end{eqnarray}
where $a$ is the lattice constant of the triangular lattice,
$r=(1-\Lambda/R)\sqrt{3}/2$ is the location of the test vortex
(scaled to $R$), $f_{vv}(x_1,x_2)=1/(x_1-x_2)-x_2^2/(x_2^2x_1-x_2)$
is the dimensionless vortex-vortex interaction, and $L=3n(n+1)+1$ is
the total number of vortices in a perfect triangular lattice with
$n$ hexagonal shells around the central vortex. The first term in
Eq. (\ref{Eq:MesoMacro}) sums up the interactions of the test vortex
with all others (with position $r_i$), using the empirically
determined fact that $\sum_{r-r_i\leq a}^i
f_{vv}(r,r_i)\approx\sum_{ka<r-r_i\leq (k+1)a}^{i}f_{vv}(r,r_i)$,
for $k=1,2,...$ ($f_{vv}$ decays as $1/\rho$ at distance $\rho$ from
the test vortex, but the number of vortices in the surrounding
shells grows proportional to $\rho$). Second term in Eq.
(\ref{Eq:MesoMacro}) describes the interaction of the test vortex
with the boundary, while the right side of the equation gives the
interaction with the shielding currents, induced by the field
$h=\frac{b+cL}{e+fL}\frac{\Phi_0}{\pi R^2}$ which stabilizes a state
with vorticity $L$ as an energy minimum (c.f. Ref. \cite{Baelus2004}
for exact coefficients). Taking $a=\Lambda\approx533$ nm to be the
limiting case for ``magnetically'' distinguishing between vortices
in the lattice, and number of shells $n=8$, Eq. (\ref{Eq:MesoMacro})
yields $R\approx 9.91\Lambda=5.28~\mu$m (and grows further with
$n$), which falls within the crossover regime where our experiments
are performed.

In conclusion, we have presented the first experimental study of
large mesoscopic superconductors (high-$T_c$ disks) which has been
able to demonstrate both macro- and mesoscopic behaviour. In disks
of 20 $\mu$m diameter, we have imaged the growth of the flux dome
and successfully fitted its observed expansion using analytic
continuum models. The latter models failed however to describe flux
penetration in 10 $\mu$m disks, in which we were able to directly
image the compression of small vortex clusters in increasing fields.
Even at higher fields where the continuum model regains validity,
the system continues to exhibit discreteness through characteristic
features in the ``local'' $M$($H_z$) curves. These measurements, in
combination with numerical simulations, allow us to identify the
magnetic signatures of individual vortices penetrating the centre of
the disk, the assembly process of vortices into different shell
formations, and the collective locking of flux lines into
configurations which are simultaneously triangular and concentric.
Finally, the competing effects between Abrikosov lattice and
shell-like ordering have allowed us to develop an analytic criterion
which places the ``large mesoscopic'' regime for superconducting
disks at radii between 10$\lambda$ and 20$\lambda$.

This work was supported by EPSRC-UK under grant No. GR/D034264/1 and
the Royal Society through an International Joint Project No.
2005/R1. M.V.M. is a Marie-Curie Intra-European Fellow at the
University of Bath. J.R.C. acknowledges support by the Department of
Energy - Basic Energy Sciences under Contract No. DE-AC02-07CH11358.
T.T. acknowledges support from JSPS bilateral cooperative program.

\end{document}